\documentclass[12pt]{article}%
\usepackage{amsmath,latexsym}
\usepackage{graphicx}
\usepackage{amsmath}
\usepackage{amsfonts}
\usepackage{amssymb}%
\begin{document}

\title{{Using embedding theorems to account for the
   extreme properties of traversable wormholes}}
   \author{
Peter K.F. Kuhfittig*\\  \footnote{kuhfitti@msoe.edu}
 \small Department of Mathematics, Milwaukee School of
Engineering,\\
\small Milwaukee, Wisconsin 53202-3109, USA}

\date{}
 \maketitle
 \noindent

\begin{abstract}\noindent
Embedding theorems, which have a long history
in the general theory of relativity, are
used in this paper to account for two of
the more troubling aspects of Morris-Thorne
wormholes, (1) the origin of exotic matter
and the amount needed to sustain a wormhole,
and (2) the enormous radial tension that
is characteristic of wormholes with
moderately-sized throats.  Attributing the
latter to exotic matter ignores the fact
that exotic matter was introduced for a
completely different reason and is usually
present in only small quantities.   \\
\end{abstract}


\section{Introduction}\label{S:introduction}
Embedding theorems have their origin in
classical geometry.  Both hyperbolic and
elliptic non-Euclidean geometry have an
intrinsic constant curvature and can be
visualized as the surface of a pseudosphere
or ordinary sphere, respectively, in
Euclidean three-space.  The two-dimensional
curved surfaces are thereby embedded in
a three-dimensional flat space.  More
generally, according to Campbell's theorem
\cite{jC26}, a Riemannian space can be
embedded in a higher-dimensional flat
space: an $n$-dimensional Riemannian space
is said to be of embedding class $m$ if
$m+n$ is the lowest dimension $d$ of the 
flat space in which the given space can be
embedded.  Here $d=\frac{1}{2}n(n-1)$.
So a four-dimensional Riemannian space
is of class two since it can be embedded
in a six-dimensional flat space.  Moreover,
a line element of class two can be reduced
to a line element of class one by a suitable
coordinate transformation \cite{sM19,
sM16, MRG17, sM17, MM17, MG17}, discussed
further in Sec. \ref{S:embedding}.

We continue now by recalling that
wormholes are handles or tunnels connecting
widely separated regions of our Universe
or different universes altogether.  Morris
and Thorne \cite{MT88} proposed the
following static and spherical symmetric
line element for a wormhole spacetime:
\begin{equation}\label{E:line1}
  ds^{2}=-e^{\nu(r)}dt^{2}+e^{\lambda(r)}dr^2
  +r^{2}(d\theta^{2}
  +\text{sin}^{2}\theta\,d\phi^{2}),
\end{equation}
where
\begin{equation}\label{E:form1}
   e^{\lambda(r)}=\frac{1}
  {1-\frac{b(r)}{r}}.
\end{equation}
(We are using units in which $c=G=1$.)  Here
$\nu=\nu(r)$ is called the \emph{redshift
function}, which must be everywhere finite
to prevent the occurrence of an event
horizon.  The function $b=b(r)$ is called
the \emph{shape function} since it
determines the spatial shape of the
wormhole when viewed, for example, in an
embedding diagram \cite{MT88}.  For the
shape function, we must have $b(r_0)=r_0$,
where the spherical surface $r=r_0$ is
called the \emph{throat} of the wormhole.
Other requirements are $b'(r_0)< 1$, called
the \emph{flare-out condition}, while
$b(r)<r$ for $r>r_0$.  A final requirement is
asymptotic flatness: $\text{lim}_{r\rightarrow
\infty}\nu(r)=0$ and $\text{lim}_{r\rightarrow
\infty}b(r)/r=0$.

The flare-out condition can only be met by
violating the null energy condition (NEC),
$T_{\alpha\beta}k^{\alpha}k^{\beta}\ge 0$,
for all null vectors $k^{\alpha}$, where
$T_{\alpha\beta}$ is the energy-momentum
tensor.  Matter that violates the NEC is
called ``exotic" in Ref. \cite{MT88}.  For
the outgoing null vector $(1,1,0,0)$, the
violation becomes
\begin{equation}\label{E:NEC1}
   T_{\alpha\beta}k^{\alpha}k^{\beta}=
   \rho +p_r<0.
\end{equation}
Here $T^t_{\phantom{tt}t}=-\rho$ is the
energy density, $T^r_{\phantom{rr}r}= p_r$
is the radial pressure, and
$T^\theta_{\phantom{\theta\theta}\theta}=
T^\phi_{\phantom{\phi\phi}\phi}=p_t$ is
the lateral (transverse) pressure.

While the need for exotic matter is rather
problematical, it is not a conceptual
problem, as we know from the Casimir
effect \cite{MT88}.  In other words,
exotic matter can be made in the
laboratory.  An open question is whether
enough could be produced to sustain a
macroscopic traversable wormhole.  It
is proposed in this paper that exotic
matter may be part of the induced-matter
theory, discussed below.

Our second problem, also to be addressed
via the embedding theory, is the enormous
radial tension at the throat.  To that
end, we first need to recall that the
radial tension $\tau(r)$ is the negative
of the radial pressure $p_r(r)$.  It is
pointed out in Ref. \cite{MT88} that the
Einstein field equations can be rearranged
to yield $\tau(r)$: reintroducing $c$ and
$G$ for now, the radial tension is given
by
\begin{equation}\label{E:tau}
   \tau(r)=\frac{b(r)/r-[r-b(r)]\nu'(r)}
   {8\pi Gc^{-4}r^2}.
\end{equation}
From this condition it follows that the
radial tension at the throat is
\begin{equation}\label{E:large}
  \tau(r_0)=\frac{1}{8\pi Gc^{-4}r_0^2}\approx
   5\times 10^{41}\frac{\text{dyn}}{\text{cm}^2}
   \left(\frac{10\,\text{m}}{r_0}\right)^2.
\end{equation}
In particular, for $r_0=3$ km, $\tau(r)$
has the same magnitude as the pressure
at the center of a massive neutron star
\cite{MT88}.  Attributing this outcome to
exotic matter ignores the fact that exotic
matter was introduced for a completely
different reason, ensuring a violation of
the NEC.  For example, it is known that
dark matter and phantom dark energy can
support traversable wormholes due to the
NEC violation but could not explain the
large radial tension. In fact, according
to Eq. (\ref{E:large}), to avoid a large
$\tau(r_0)$, the throat radius $r=r_0$
would have to be extremely large, so
that such a wormhole could only exist
on a very large scale.

An even more serious problem is the claim
made in Ref. \cite{VKD}: the total amount
of exotic matter required can be infinitely
small.  If so, then the exotic matter
cannot  be responsible for the large
radial tension.

The question of an arbitrarily small 
violation of the null energy condition 
is also discussed in Ref. \cite{oZ07}.  
It is shown that whenever the amount of 
exotic matter tends to zero, then either 
the resulting structure is not a wormhole 
or the stresses developing at the throat 
become infinite.

The aforementioned wormholes requiring
large throat sizes for their existence
have remained topics of interest in
cosmology, however, as illustrated in
Refs. \cite{fR14a, fR14b}.  In a similar
vein, Ref. \cite{OH16} discusses traversable
wormholes in spherical stellar systems.
(See also Ref. \cite{OJS19}.)  Another
area of interest in a dark-matter medium
is the possible detection of wormholes
by means of gravitational lensing.  The
determination of the deflection angles
of black holes and wormholes is discussed
in Refs. \cite{aO19, aO20}.

Before continuing, let us list the Einstein
field equations, referring to line
element (\ref{E:line1}):

\begin{equation}\label{E:Einstein1}
8\pi \rho=e^{-\lambda}
\left[\frac{\lambda^\prime}{r} - \frac{1}{r^2}
\right]+\frac{1}{r^2},
\end{equation}
\begin{equation}\label{E:Einstein2}
8\pi p_r=e^{-\lambda}
\left[\frac{1}{r^2}+\frac{\nu^\prime}{r}\right]
-\frac{1}{r^2},
\end{equation}
and
\begin{equation}\label{E:Einstein3}
8\pi p_t=
\frac{1}{2} e^{-\lambda} \left[\frac{1}{2}(\nu^\prime)^2+
\nu^{\prime\prime} -\frac{1}{2}\lambda^\prime\nu^\prime +
\frac{1}{r}({\nu^\prime- \lambda^\prime})\right].
\end{equation}

\section{The role of embedding}
    \label{S:embedding}
Embedding theorems have a long history
in the general theory of relativity.  In
particular, according to Refs. \cite{WP92,
SW03}, the vacuum field equations in
five dimensions yield the Einstein field
equations \emph{with matter}, called the
induced-matter theory.  What we perceive
as matter, is simply the impingement of
the higher-dimensional space onto ours.
Moreover, it is noted in the Introduction
that a metric of class two can be reduced
to a metric of class one  and can therefore
be embedded in the five-dimensional flat
spacetime
\begin{equation}\label{E:line3}
   ds^2=-(dz^1)^2+(dz^2)^2+(dz^3)^2+
   (dz^4)^2+(dz^5)^2;
\end{equation}
the coordinate transformation is
$z^1=\sqrt{K}\,e^{\nu/2}\,\text{sinh}
\frac{t}{\sqrt{K}}$, $z^2=
\sqrt{K}\,e^{\nu/2}\,\text{cosh}
\frac{t}{\sqrt{K}}$,
$z^3=r\,\text{sin}\,\theta\,
\text{cos}\,\phi$, $z^4=r\,\text{sin}
\,\theta\, \text{sin}\,\phi$, and
$z^5=r\,\text{cos}\,\theta$.  As a
result,
\begin{equation}\label{E:line2}
ds^{2}=-e^{\nu}dt^{2}
 +\left[1+\frac{1}{4}Ke^{\nu}(\nu')^2\right]dr^2
+r^{2}(d\theta^{2}+\text{sin}^{2}\theta\,
d\phi^{2}).
\end{equation}
Metric (\ref{E:line2}) is equivalent to
metric (\ref{E:line1}) if
\begin{equation}\label{E:lambda}
   e^{\lambda}=1+\frac{1}{4}Ke^{\nu}(\nu')^2,
\end{equation}
where $K>0$ is a free parameter.  The
result is a metric of embedding class
one.  Eq. {(\ref{E:lambda}) can also be
obtained from the Karmarkar condition
\cite{kK48}
\begin{equation*}
  R_{1414}=
  \frac{R_{1212}R_{3434}+R_{1224}R_{1334}}
  {R_{2323}},\quad R_{2323}\neq 0.
\end{equation*}
In fact, Eq. (\ref{E:lambda}) is a solution
of the differential equation
\begin{equation*}
   \frac{\nu'\lambda'}{1-e^{\lambda}}=
   \nu'\lambda'-2\nu''-(\nu')^2,
\end{equation*}
which is readily solved by separation
of variables. So $K$ is actually an
arbitrary constant of integration.

\section{The wormhole solution}
To help make our wormhole solution
physically acceptable, we will use the
following redshift function proposed by
Lake \cite{kL03}:
\begin{equation}\label{E:nu}
   \nu(r)=n\,\text{ln}\,(1+Ar^2),\quad
   n\geq 1,
\end{equation}
where $A$ is a positive constant.
According to Ref. \cite{kL03}, this
class of monotone increasing functions
generates all regular static spherically
symmetric perfect-fluid solutions of
the Einstein field equations.  Eq.
(\ref{E:nu}) has proved to be extremely
useful in the study of compact stellar
objects \cite{sM19, sM16}.  While Eq.
(\ref{E:nu}) can be written $e^{\nu}
=(1+Ar^2)^n$, it can also be generalized
to $e^{\nu}=B(1+Ar^2)^n$, $B>0$, since
the resulting $\nu$ is still a monotone
increasing function.  For our purposes,
however, it is sufficient to let $n=1$.
The resulting form is
\begin{equation}\label{E:redshift}
   e^{\nu}=B(1+Ar^2),\quad A,B>0.
\end{equation}
While $A$ is still a free parameter, the
constant $B$ can be determined from the
junction conditions discussed at the
end of the section.

To see how the free parameter $K$ comes
into play, let us determine $b(r)$ from
Eq. (\ref{E:lambda}) by inspection:
\begin{equation}\label{E:shape}
   b(r)=r\left[1-\frac{1}
   {1+\frac{1}{4}Ke^{\nu(r)}[\nu'(r)]^2}
   \right]+\frac{r_0}{1+\frac{1}
   {4}Ke^{\nu(r_0)}[\nu'(r_0)]^2};
\end{equation}
so $b(r_0)=r_0$.

To check the flare-out condition, we
start with
\begin{equation}\label{E:bprime}
   b'(r)=1-\frac{1}{1+\frac{1}
   {4}Ke^{\nu}(\nu')^2}
   +r\,\frac{\frac{1}{4}Ke^{\nu}\nu'
   [(\nu')^2+2\nu'']}{\left[1+\frac{1}
   {4}Ke^{\nu}(\nu')^2\right]^2}.
\end{equation}
The last term on the right can be
rewritten as
\begin{equation*}
   \frac{r}{\nu'}\,
   \frac{ \frac{1}{4}Ke^{\nu}(\nu')^2
   [(\nu')^2+2\nu'']}
   {[1+ \frac{1}{4}Ke^{\nu}(\nu')^2]
   [1+ \frac{1}{4}Ke^{\nu}(\nu')^2]}.
\end{equation*}
Since the free parameter $K$ can be
arbitrarily large, the 1 in the
expression $1+ \frac{1}{4}Ke^{\nu}(\nu')^2$
can be neglected, yielding
\begin{equation*}
   \frac{r}{\nu'}\,\frac{(\nu')^2+2\nu''}
   {1+ \frac{1}{4}Ke^{\nu}(\nu')^2}.
\end{equation*}
From Eq. (\ref{E:redshift}), we also
obtain
\begin{equation}\label{E:derivatives}
   \nu'=\frac{2Ar}{1+Ar^2} \quad
   \text{and} \quad \nu''=
   \frac{2A-2A^2r^2}{(1+Ar^2)^2}.
\end{equation}
Substituting in Eq. (\ref{E:bprime}), we
have at the throat,
\begin{equation}\label{E:bprime2}
   b'(r_0)=1+\frac{\frac{1-Ar_0^2}
      {1+Ar_0^2}}{1+\frac{1}{4}K
      e^{\nu(r_0)}[\nu'(r_0)]^2}
      =1+\frac{1-Ar_0^2}
      {1+Ar_0^2+KBA^2r_0^2}.
\end{equation}
Since $K$ can be made as large as we
please,
\begin{equation}\label{E:flair}
   b'(r_0)<1 \quad \text{for} \quad
   r_0>\frac{!}{\sqrt{A}}.
\end{equation}
So the flare-out condition is
satisfied whenever $r_0>1/\sqrt{A}$.

Our final topic in this section is
asymptotic flatness.  Since
$\text{lim}_{r\rightarrow
\infty}\nu'(r)=0$ from Eq.
(\ref{E:derivatives}), it follows
from Eq. (\ref{E:shape}) that
$\text{lim}_{r\rightarrow
\infty}b(r)/r=0$.  Unfortunately, it is
not true that $\text{lim}_{r\rightarrow
\infty}\nu(r)=0$.  The wormhole
spacetime must therefore be cut off
at some $r=a$ and joined to the
exterior Schwarzschild solution
\begin{equation}
  ds^{2}=-\left(1-\frac{2M}{r}\right)dt^{2}
  +\frac{dr^2}{1-\frac{2M}{r}}
  +r^{2}(d\theta^{2}
  +\text{sin}^{2}\theta\,d\phi^{2}).
\end{equation}
So
\[e^{\nu(a)}=B(1+Aa^2)=1-\frac{2M}{a}
\]
and
\begin{equation}
   B=\frac{1-\frac{2M}{a}}{1+Aa^2}.
\end{equation}
(We will continue to use $B$ in place
of its actual value.)  The junction at
$r=a$ also yields the mass of the
wormhole: $M=\frac{1}{2}b(a)$.

\section{The high radial tension and the
amount of exotic matter}
\subsection{The high radial tension}
Accounting for the high radial tension is
one of the goals in this paper.  As noted
in the Introduction, attributing this
property to exotic matter ignores the fact
that exotic matter was introduced to
ensure a violation of the NEC.  The reason
for the high radial tension must therefore
be sought elsewhere.

Using Eq. (\ref{E:redshift}) in Eq.
(\ref{E:Einstein2}), we obtain the
following expression for the radial pressure:
\begin{equation}\label{E:pressure}
   p_r(r)=\frac{A(2-KBA)}{8\pi(1+Ar^2
      +KBA^2r^2)}.
\end{equation}
Recall that we are primarily interested in
relatively small throat sizes, since
$\tau(r_0)$ is small for large values of
$r_0$.  Now consider
\begin{equation}\label{E:limittau}
  \text{lim}_{r_0\rightarrow 0} \tau(r_0)=
  \text{lim}_{r_0\rightarrow 0}(-p_r(r_0))=
  \text{lim}_{r_0\rightarrow 0}\frac
  {A(-2+KBA)}{8\pi(1+Ar_0^2+KBA^2r_0^2)}
  =\frac{A}{8\pi}(-2+KBA).
\end{equation}
So $\tau(r_0)$ can be made as large as
required due to the free parameter $K$,
as long as $r_0$ is relatively small.

An alternative to the present theory is
discussed in Ref. \cite{pK20}.  It is
shown that the large radial tension can
be accounted for via noncommutative
geometry, an offshoot of string theory,
or by the existence of a small extra
spatial dimension.

\subsection{The amount of exotic matter}
As noted earlier, since the vacuum field
equations in five dimensions yield the
Einstein field equations \emph{with
matter}, this may very well include
exotic matter.  So the amount may seem
irrelevant.  Given its problematical
nature, however, it is desirable to
keep the amount to a minimum.  The
actual amount required to sustain a
traversable wormhole was first considered
by Visser et al. \cite{VKD} and then
extended by Nandi et al. \cite{NZK}:
\begin{equation}\label{E:Nandi}
   \Omega =\int^{2\pi}_0\int^{\pi}_0
   \int^{\infty}_{r_0}(\rho +p_r)
   \sqrt{-g}\,\,drd\theta d\phi.
\end{equation}

Since $p_r(r)$ is negative and large in
absolute value, $\rho+p_r$ is also
negative, as expected, since the flare-out
condition has been met.  But we still need
to show that $\rho+p_r$ can be small in
absolute value.  From Eqs. (\ref{E:Einstein1})
and (\ref{E:pressure}),
\begin{equation}
   8\pi(\rho+p_r)=\frac{1}
   {1+\frac{1}{4}Ke^{\nu}(\nu')^2}\frac{1}{r^2}
   \left[\frac{r}{\nu'}\,[(\nu')^2+2\nu'']-1\right]
   +\frac{1}{r^2}+\frac{A(2-KBA)}
   {1+Ar^2+KBA^2r^2},
\end{equation}
which is indeed negative for small $r_0$ by
Eq. (\ref{E:limittau}).  To see the full
effect of the free parameter $K$, we simply
observe that
\[
    \text{lim}_{K\rightarrow \infty}
    \frac{A(2-KBA)}{1+Ar^2+KBA^2r^2}
    =\frac{-BA^2}{BA^2r^2}=-\frac{1}{r^2}
\]by L'Hospital's rule.  So
\begin{equation}
    \text{lim}_{K\rightarrow \infty}
    8\pi(\rho+p_r)=0.
\end{equation}
Since $\rho+p_r$ can be made small thanks
to the embedding theory, it follows from
Eq. (\ref{E:Nandi}) that the total amount
of exotic matter can also be small.

\section{Conclusions}
An $n$-dimensional Riemannian space is said
to be of embedding class $m$ if $m+n$ is
the lowest dimension of the flat space in
which the given space can be embedded.  This
paper deals with wormholes in spacetimes
of embedding class one.  The goal is to
use the embedding theory to account for
two of the most troubling aspects of wormhole
physics, the need for exotic matter and the
enormous radial tension at the throat of any
moderately-sized wormhole.

It is proposed in this paper that the
existence of exotic matter can be attributed
to the higher-dimensional embedding space,
thereby becoming part of the induced-matter
theory.  So while the amount of exotic matter
may therefore seem irrelevant, its
problematical nature demands that the total
amount be kept to a minimum.  This requirement
can be met thanks to the free parameter $K$.

The large radial tension at the throat is
usually attributed to the presence of
exotic matter.  The problem is that exotic
matter was introduced for a totally
different reason, the need for violating
the NEC.  Reducing the amount makes the
large radial tension even harder to
explain.  If the amount is infinitely
small \cite{VKD}, this explanation breaks
down entirely.  The embedding theory has
proved to be an effective way to account
for the high radial tension without
relying on exotic matter.
\\
\\
\noindent
CONFLICTS OF INTEREST
\\
\noindent
The author declares that there are no
conflicts of interest regarding the
publication of this paper.

\end{document}